\documentclass{elsart}

\usepackage{graphicx}

\usepackage{amssymb}

\begin{document}

\begin{frontmatter}



\title{Electromagnetic Probes of Strongly Interacting Matter}


\author{J. Kapusta}

\address{School of Physics and Astronomy, University of Minnesota\\
 Minneapolis, MN 55455, USA}

\begin{abstract}
Photons and dileptons are being used to probe the properties of nuclear and
quark-gluon matter at high energy densities.  This is an area where theory and
experiment are driving each other to obtain solid results.  However, it is
important to clearly separate the assumptions and conclusions concerning the
correlation and response functions of a system in thermal equilibrium from the
space-time dynamics used to model the evolution of the matter created in the
nuclear collision.
\end{abstract}


\end{frontmatter}

{\bf 1 Introduction}

It has been thirty years since Feinberg suggested that soft quarks, anti-quarks
and gluons could interact with each other prior to hadronization to produce
photons and dileptons of moderate energy.  This idea was soon
developed by Shuryak and others.  (For a brief history see the
annonated reprint volume by M\"uller, Rafelski and me \cite{kapusta1}.)  It has
been seventeen years since the DLS (DiLepton Spectrometer) Collaboration
published their first measurements of dilepton emission from nuclear collisions
in the GeV per nucleon beam energy regime \cite{DLS}.  Much work has been done
by theorists and experimentalists since these pioneering efforts.  Here I would
like to discuss some of the accomplishments and some of the important issues in
this field.  It should also serve to introduce the following articles.

The mean free path for real or virtual photons in hot and dense matter is very
large, typically more than $10^2$ to $10^4$ fm.  This is due to the relative
smallness of the fine structure constant.  It makes these good probes of the
medium because they do not suffer final state interactions, and therefore convey
information about the system directly to the detectors.  The penalty is that the
production rate is small, and the background from hadronic decays is large.

What is the most that we can learn from electromagnetic probes?\\
$\bullet$ We can infer the electromagnetic current-current correlation function
in the medium {\it if we know the dynamical evolution of the system}.\\
$\bullet$ We can infer the dynamical evolution of the system {\it if we know the
electromagnetic current-current correlation function in the medium}.\\
One must decide which one of these is the goal.  One should not mix them up.
Undoubtedly there are a continuously infinite number of ways to parameterize the
electromagnetic emission rate and the dynamical evolution of the nuclear
collision in such a way as to reproduce the data. Fortunately there are
constraints on theoretical calculations of the emission rates, and there are the
full set of single and multi-particle hadronic spectra that must be reproduced
by the theory in addition to the photons and dileptons.  Success cannot be
claimed unless detailed cross comparisons are made in all respects.

{\bf 2 Electromagnetic emission rates}

The formal expressions for the electromagnetic emission rates in relativistic
quantum field theory were worked out by various people \cite{EMrate}.  For
photons the rate is
\begin{equation}
\omega \frac{d^3 R }{d^3 k} = - \frac{g^{\mu \nu}}{(2 \pi)^3} {\rm Im}
\Pi^R_{\mu
\nu} (\omega, {\bf k}) \frac{1}{e^{\beta \omega} - 1} \, ,
\end{equation}
and for lepton pairs it is
\begin{equation}
E_+ E_- \frac{d^6 R}{d^3 p_+ d^3 p_-}=\frac{2 e^2}{(2 \pi)^6} \frac{1}{k^4}
L^{\mu\nu}(p_+, p_-) {\rm Im}
\Pi^R_{\mu \nu} (\omega, {\bf k}) \frac{1}{e^{\beta \omega} - 1} \, .
\end{equation}
Here $R$ is the rate (number per unit time per unit volume), $\Pi$ is the photon
self-energy in the thermal medium, and $L$ is a kinematic tensor involving the
four-momenta of the leptons.  The electromagnetic spectra will be direct probes
of the in-medium photon self-energy or current-current correlation function if
we have a dynamical evolution scenario over which to integrate the rates.

A very useful theoretical approach to the dilepton mass range from a few hundred
MeV to just above a GeV is vector-meson dominance.  The current-field identity
of Sakurai \cite{Sakurai} expresses the electromagnetic current in terms of the
vector-meson fields.
\begin{equation}
J_\mu = - \frac{e}{g_\rho} m_\rho^2 \rho_\mu - \frac{e}{g_\omega} m_\omega^2
\omega_\mu - \frac{e}{g_\phi} m_\phi^2 \phi_\mu
\end{equation}
Considering just the $\rho$-meson, we have
Im$\langle \rho^{\mu}\rho^{\nu}\rangle$ $\rightarrow$
Im$\langle D_{\rho}^{\mu\nu}\rangle$ $\rightarrow$
Im$\langle J^{\mu}J^{\nu}\rangle$ $\rightarrow$
Im$\langle \Pi^{\mu\nu}\rangle$.  This imaginary part is readily expressed in
terms of the spectral density in the medium, a quantity of fundamental interest.
The $\rho$-meson propagator is expressed in terms of two scalar self-energies,
$F$ and $G$.
\begin{equation}
D_{\rho}^{\mu \nu} = - \frac{P^{\mu \nu}_L}{k^2 - m_\pi^2 - F} - \frac{P^{\mu
\nu}_T}{k^2 - m_\rho^2 - G} - \frac{k^\mu k^\nu}{m_\rho^2 k^2}
\end{equation}
Of course, the $\omega$, $\phi$, and $J/\psi$ vector-mesons need to be included
too.  Generally one would expect a peak at each of the corresponding masses,
perhaps shifted up or down and broadened relative to the vacuum.  Hence the
spectral densities in the medium shape the observed spectra.

There are constraints on the spectral densities arising from the Weinberg sum
rules \cite{Weinberg} generalized to finite temperature \cite{EdJoe} in the
limit of exact chiral symmetry.  They are
\begin{eqnarray}
\int_0^{\infty} \frac{d\omega \, \omega}{\omega^2-{\bf p}^2}
\left[ \rho_V^L(\omega,{\bf p})
-\rho_A^L(\omega,{\bf p}) \right] & = & 0; \;\;\;
\int_0^{\infty} d\omega \, \omega
\left[ \rho_V^L(\omega,{\bf p})
-\rho_A^L(\omega,{\bf p}) \right]  =  0 \nonumber \\
\int_0^{\infty} d\omega \, \omega
\left[ \rho_V^T(\omega,{\bf p})
-\rho_A^T(\omega,{\bf p}) \right] & = & 0
\end{eqnarray}
where the subscripts $V$ and $A$ refer to vector and axial-vector while the
superscripts $L$ and $T$ refer to longitudinal and transverse.  The pion couples
to the longitudinal part of the axial-vector current.  As the critical
temperature is approached, this coupling goes to zero, which is equivalent to
$f_{\pi}(T \rightarrow T_c) \rightarrow 0$.  There are a variety of
possibilities for satisfying these sum rules as the temperature increases to
$T_c$.\\
$\bullet$ The spectral densities mix (Dey-Eletsky-Ioffe mixing
\cite{EletskyIoffe}).\\
$\bullet$ The $\rho$ and $a_1$ masses become degenerate: both go up, both go
down (Brown-Rho scaling \cite{BrownRho}), or one goes up and the other goes
down).\\
$\bullet$ The widths become so large that the vector and axial-vector mesons
melt away.\\
Of course reality may be a combination of all of the above.  Unfortunately
measurements of photons and dileptons alone cannot be used to investigate these
sum rules because those measurements only probe the vector-current, not the
axial-vector current.  (The latter would be probed by neutrinos.)  Nevertheless
theoretical models should obey these sum rules and any model should be tested
against them.

{\bf 3 Dileptons}

A big discovery by CERES \cite{voi,len} was the observation of an enhancement
above the hadronic decay cocktail in the mass range 300 to 700 MeV and no
obvious peak near the $\rho$ and $\omega$ mesons in high multiplicity collisions
of Pb+Au at the CERN SPS.  See the left panel of Fig. 1.  This lead to two
competing explanations: the $rho$ meson was greatly broadened in the expanding
hot and dense medium or its mass decreased with increasing energy density.  The
first explanation was studied by Rapp, Chanfray and Wambach \cite{rapp} while
the second explanation was presented by Brown and Rho \cite{BrownRho}.  Rapp
{\it et al.} calculated the modification of the $\rho$-meson propagator due to
interactions with baryons and mesons which themselves were modified by the
medium.  Brown and Rho espoused a scaling of hadron masses as powers of the
baryon density based on QCD sum rules and also on the QCD trace anomaly.
Subsequently Eletsky {\it et al.} \cite{eletsky} computed the $\rho$-meson self-
energy using the vacuum scattering amplitudes from various hadrons $h$ in the
medium using the standard formula
\begin{equation}
\Pi_{\rho}(E,p) = - 4\pi \sum_h \int \frac{d^3k}{(2\pi)^3} \,
n_h(\omega) \, \frac{\sqrt{s}}{\omega}
 \, f_{\rho h}^{(\rm cm)}(s) \, .
\end{equation}
The scattering amplitudes may be constructed essentially from experimental data
such as resonance masses and widths, phase shifts where available, and Regge
phenomenology at higher energy.  The approaches of Rapp {\it et al.} and Eletsky
{\it et al.} agree reasonably well, as shown in the right panel of Fig. 1, as
they ought to since they are both based on similar hadronic measureables and
parameters.  The main effect is due to baryon density, much less to temperature.
Neither approach sees a significant shift in the peak of the spectral density.

\begin{figure}[b]
\centering
\resizebox{6.1cm}{!}{\includegraphics{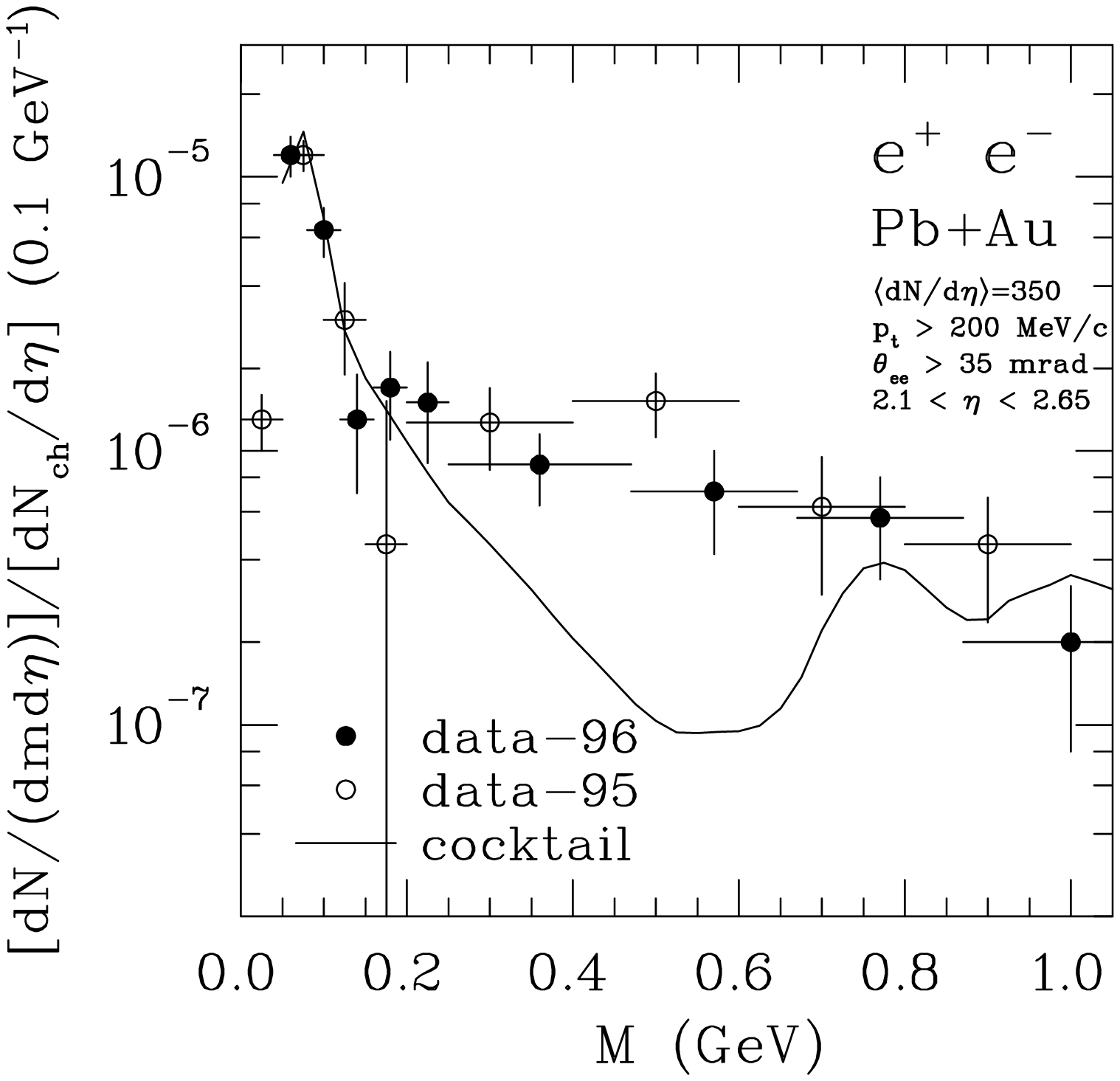}}
\hspace*{5mm}
\resizebox{6.3cm}{!}{\includegraphics{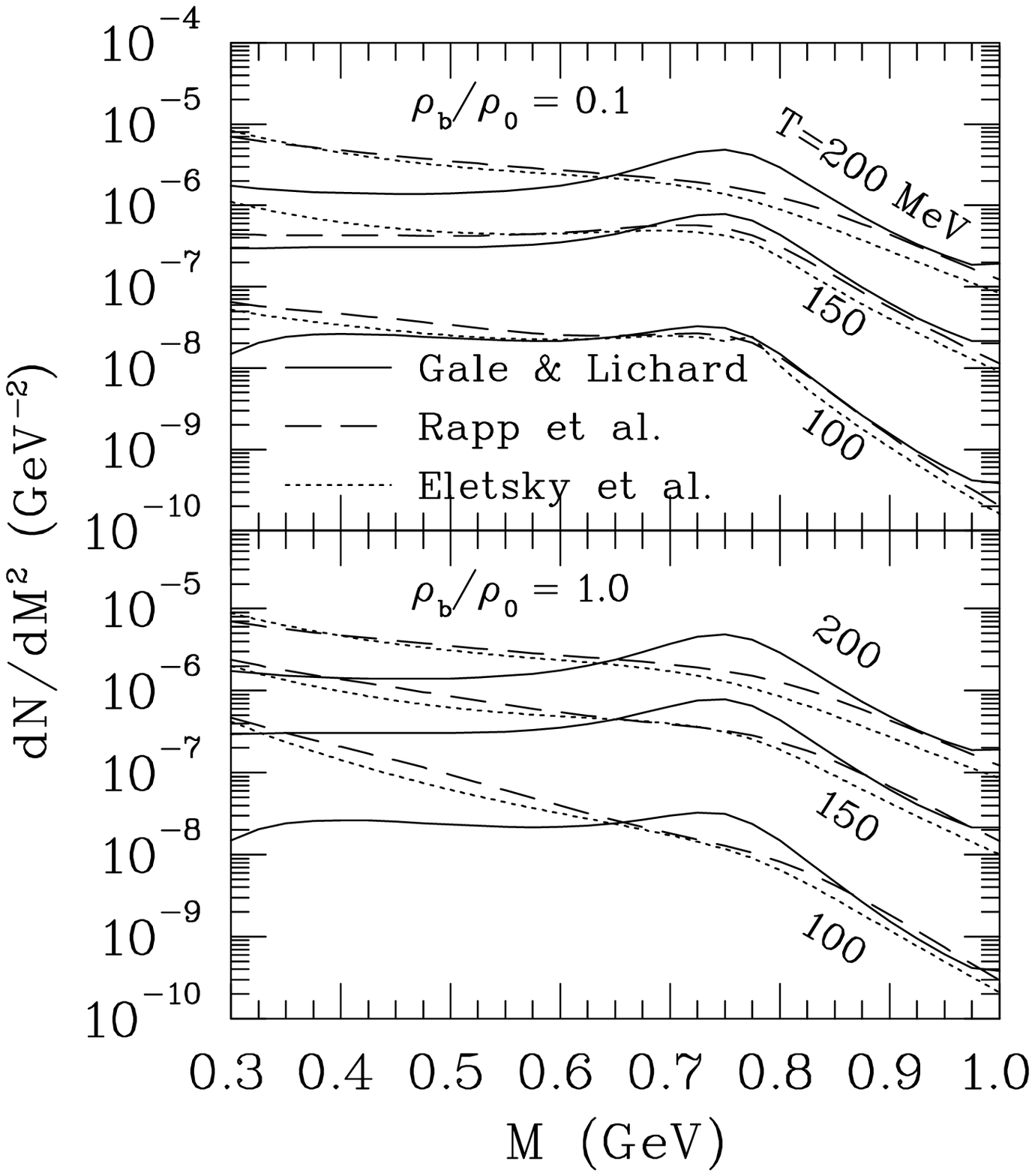}}
\caption{{\sl Left panel:} Comparison of the dilepton data for Pb-Au collisions
at 158 A GeV ('95 data Ref. \protect\cite{voi}, '96 data Ref.
\protect\cite{len}) with the contribution from the decay of hadrons after
freezeout. {\sl Right panel:} Thermal dilepton emission rates computed by Rapp
{\it et al.} \protect\cite{rapp}, Eletsky {\it et al.} \protect\cite{eletsky}
and Gale and Lichard (which has no medium effects) \protect\cite{galel}, at
various temperatures.  The baryon densities are fixed at 1/10 and 1 times the
equilibrium density of cold nuclear matter.}
\end{figure}

The rates must be folded with a dynamical evolution model to compare with the
data.  Folding with a relatively simple model shows that the hadronic decay
cocktail plus annihilation of pions as in vacuum cannot describe the data as
shown in the left panel of Fig. 2.  The $\rho$-broadening explanation as
computed by Rapp {\it et al.} does represent the data fairly well, as does the
dropping $\rho$-mass description.  The rates as computed by Eletsky {\it et al.}
were compared to the data using two different dynamical models: relativistic
hydrodynamics with parameters chosen to reproduce the hadronic spectra, and
UrQMD coarse-grained to provide contour profiles of temperature, chemical
potential, and flow velocity.  Those calculations represent the data too as
shown in the right panel of Fig. 2.

\begin{figure}[b]
\begin{center}
\resizebox{6.9cm}{!}{\includegraphics{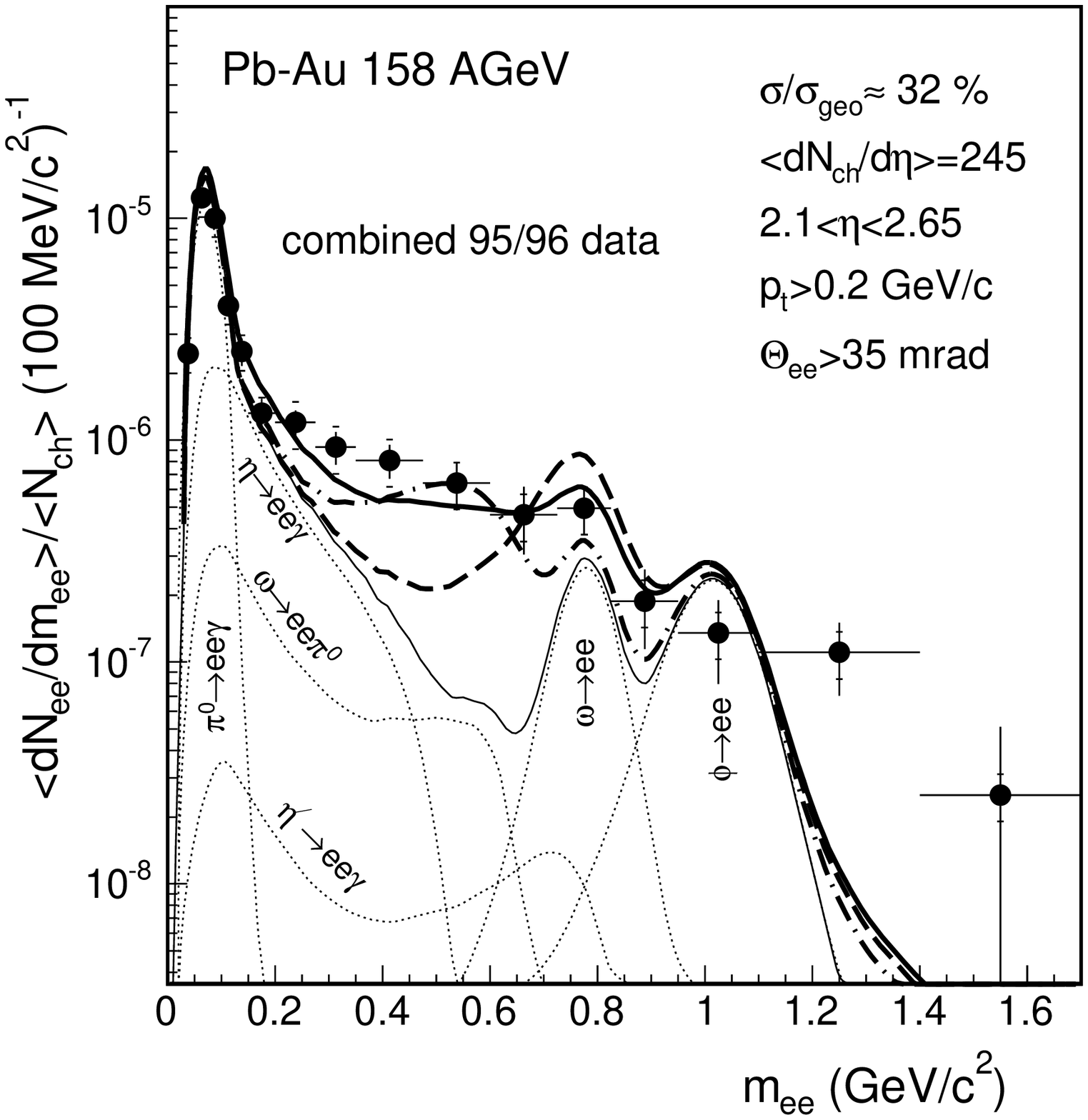}}
\hspace*{5mm}
\resizebox{5.7cm}{!}{\includegraphics{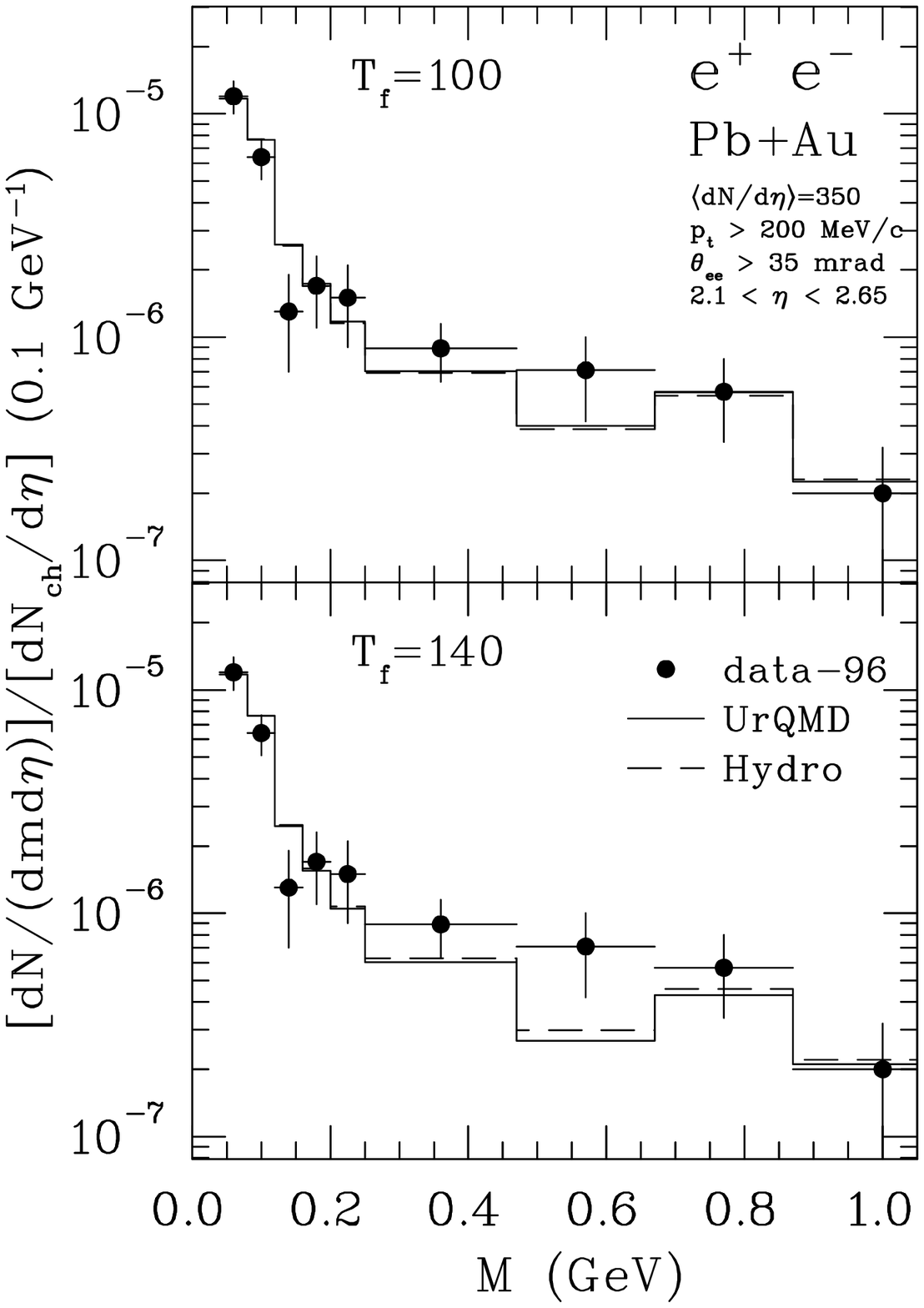}}
\caption{{\sl Left panel:} Comparison of the CERES/NA45 data with the hadron
decay cocktail, and with the vacuum $\pi\pi$ annihilations (thick dashed curve),
the medium $\pi\pi$ annihilations (thick solid curve), and with a dropping
$\rho$ mass (thick dot-dashed curve).  {\sl Right panel:} Comparison
of the dilepton data \protect\cite{len} with binned predictions of the UrQMD
model and the hydrodynamic model at two freeze-out temperatures \cite{us}.}
\end{center}
\end{figure}

New measurements on semi-central In-In collisions at the CERN SPS by NA60
\cite{NA60} provide much more data; see Fig. 3 and the paper by S. Damjanovic in
this volume.  There is sufficient statistics to allow binning in transverse
momentum of the pair.  This data does seem to allow for a clear distinction
between the two scenarios with the dropping $\rho$-mass scenario apparently
inconsistent with the data.  However, this needs to be studied more with more
sophisticated models for the dynamical evolution of the nuclear collisions.

\begin{figure}[h]
\begin{center}
\resizebox{6.2cm}{!}{\includegraphics{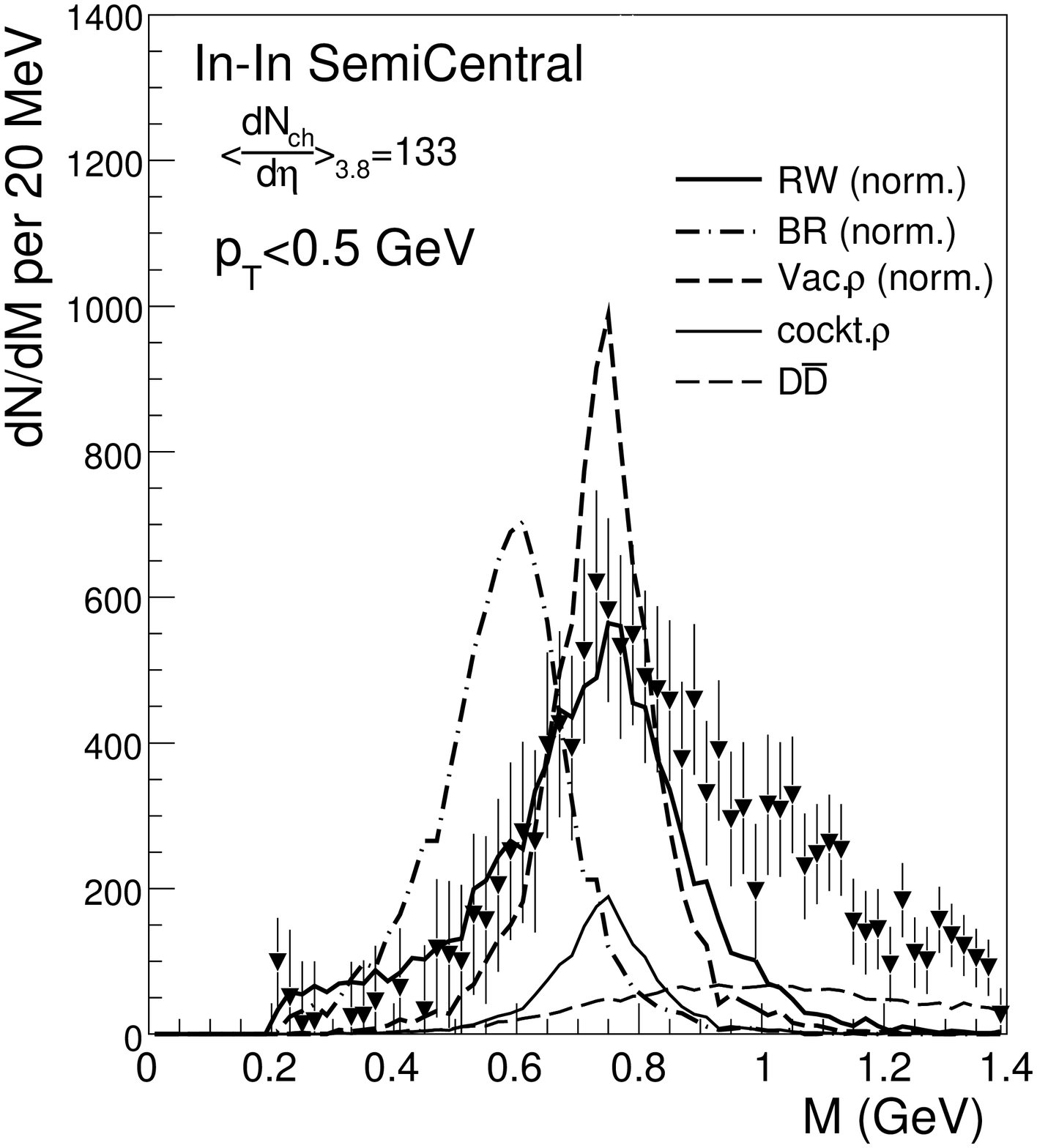}}
\hspace*{3mm}
\resizebox{6.2cm}{!}{\includegraphics{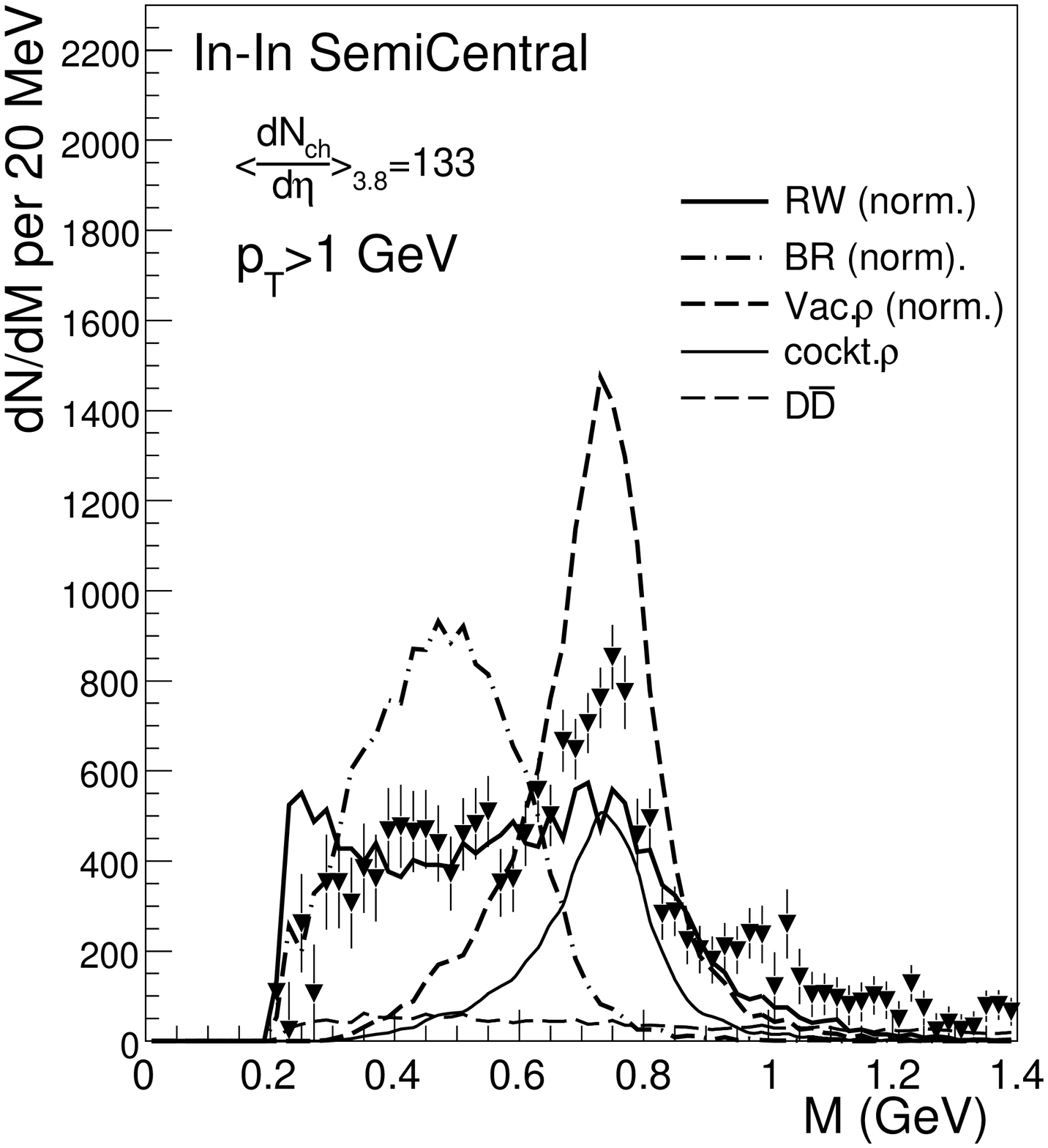}}
\caption{Data from semi-central collisions of In-In at 160 GeV per nucleon from
NA60 \cite{NA60} compared to the hadronic decay cocktail, D-meson decay, and the
spectral densities from vacuum $\rho$-meson, Rapp and Wambach, and Brown and Rho
scaling.  {\sl Left panel:} Low $p_T$.  {\sl Right panel:} High $p_T$.}
\end{center}
\end{figure}

There are several interesting theoretical approaches I have not touched on here,
including inferring spectral densities from lattice QCD (see the paper by S.
Gupta in this volume) and from the AdS/CFT correspondence (see the paper by
Kovtun in this volume).

{\bf 4 Photons}

Photons are also interesting probes of hot dense matter.  However, compared to
dileptons they appear to be a more restrictive probe since they are
characterized by their momentum whereas the dileptons also have their invariant
mass as a variable.  A soft photon in one frame of reference can be hard in
another frame, whereas a large invariant mass dilepton is hard in any frame.
However, the absolute rate for photons is larger because the thermal is
proportional to $\alpha \alpha_s$ whereas for lepton pairs it is of order
$\alpha^2 \alpha_s$.

The thermal rate for high energy photons can be computed in the QCD plasma phase
using perturbation theory and kinetic theory. It diverges logarithmically as the
quark mass goes to zero.  An infinite number of diagrams must be summed which
goes under the name "hard thermal loops".  When the photon energy $E$ is large
compared to the temperature $T$ the rate is proportional to $\ln [ E
T/(gT)^2 ]$ as computed in ref. \cite{Kap,Baier1}.  More recently, Arnold
{\it et al.} \cite{Arnold} have computed the rate when $E$ is comparable to $T$,
a much more involved calculation.  This includes the Landau-Migdal-Pomeranchuk
effect.  Therefore the thermal rate in the QCD plasma phase is relatively well
under control.  A similar statement may be made for the hadronic phase where
calculations are based on kinetic theory for scattering and annihilation of
hadrons; see the paper by Gale in this volume.

\begin{figure}[t]
\begin{center}
\resizebox{6cm}{!}{\includegraphics{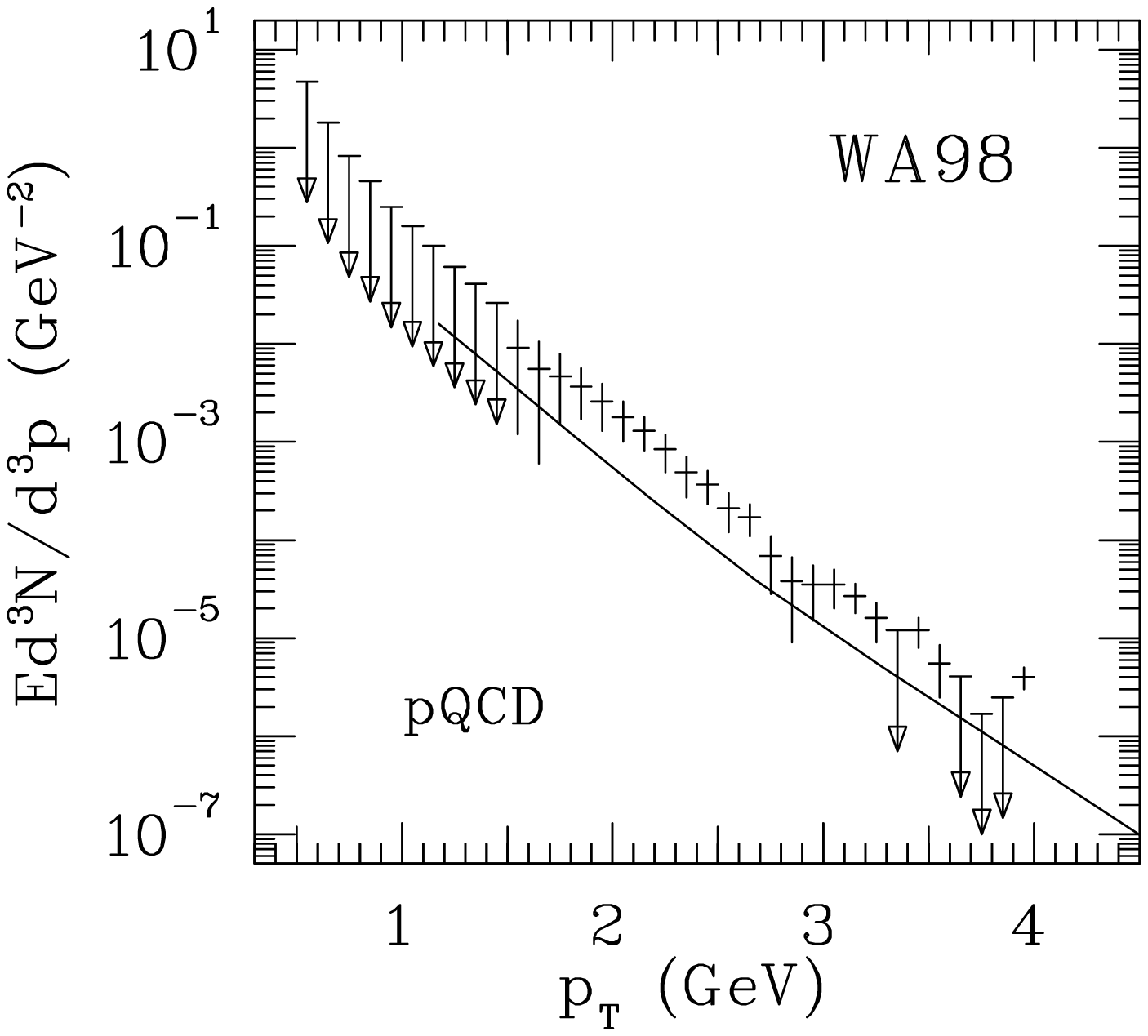}}
\hspace*{5mm}
\resizebox{6cm}{!}{\includegraphics{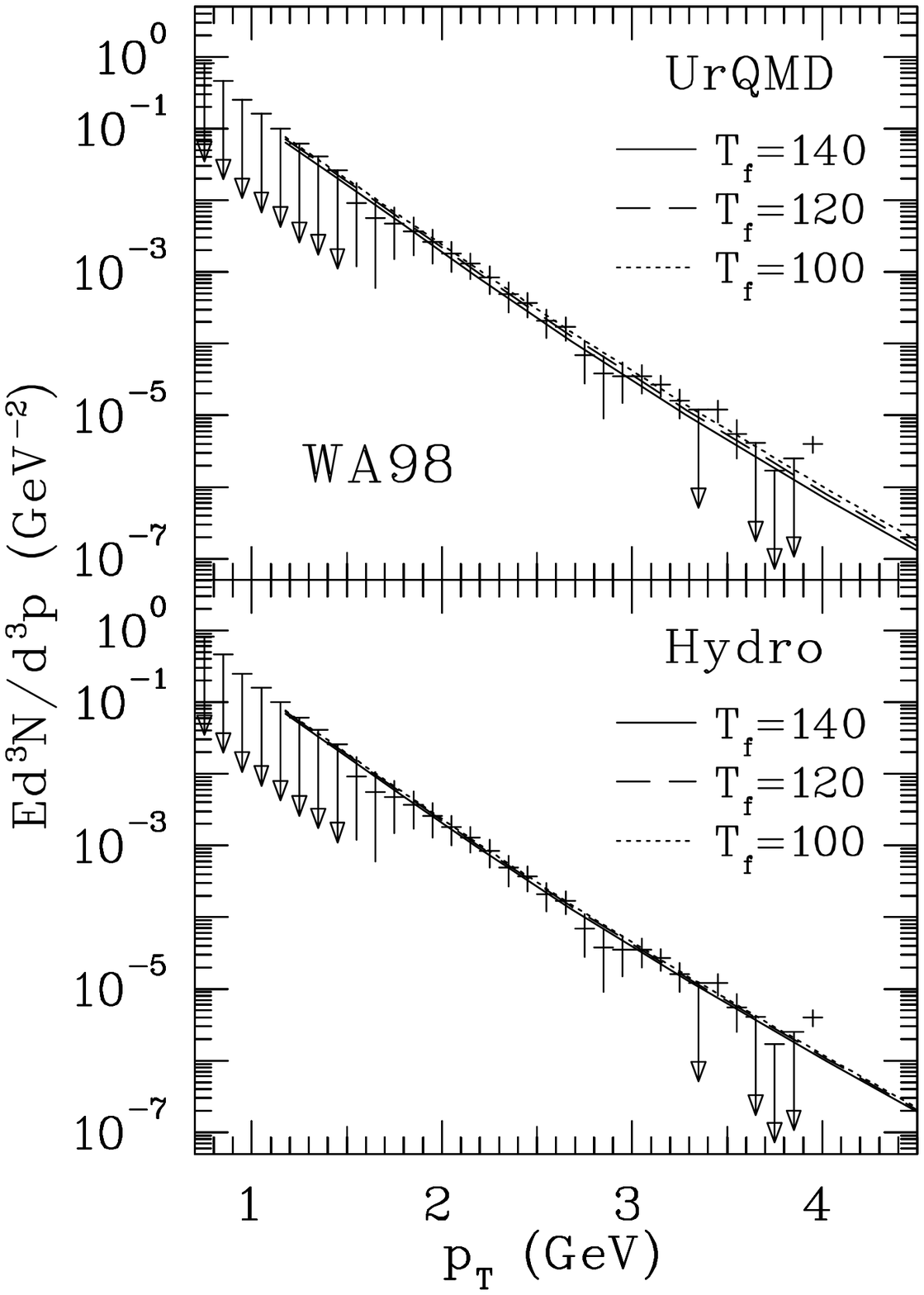}}
\caption{{\sl Left panel:} Photon spectrum from Pb-Pb collisions at 158 A GeV by
the WA98 collaboration \protect\cite{wa98} compared to a perturbative QCD
calculation.  {\sl Right panel:} Comparison of the WA98 photon spectrum
to the predictions of the UrQMD model and the hydrodynamic model at several
freeze-out temperatures \protect\cite{us}.}
\end{center}
\end{figure}

\begin{figure}[b]
\begin{center}\resizebox{6.7cm}{!}{\includegraphics{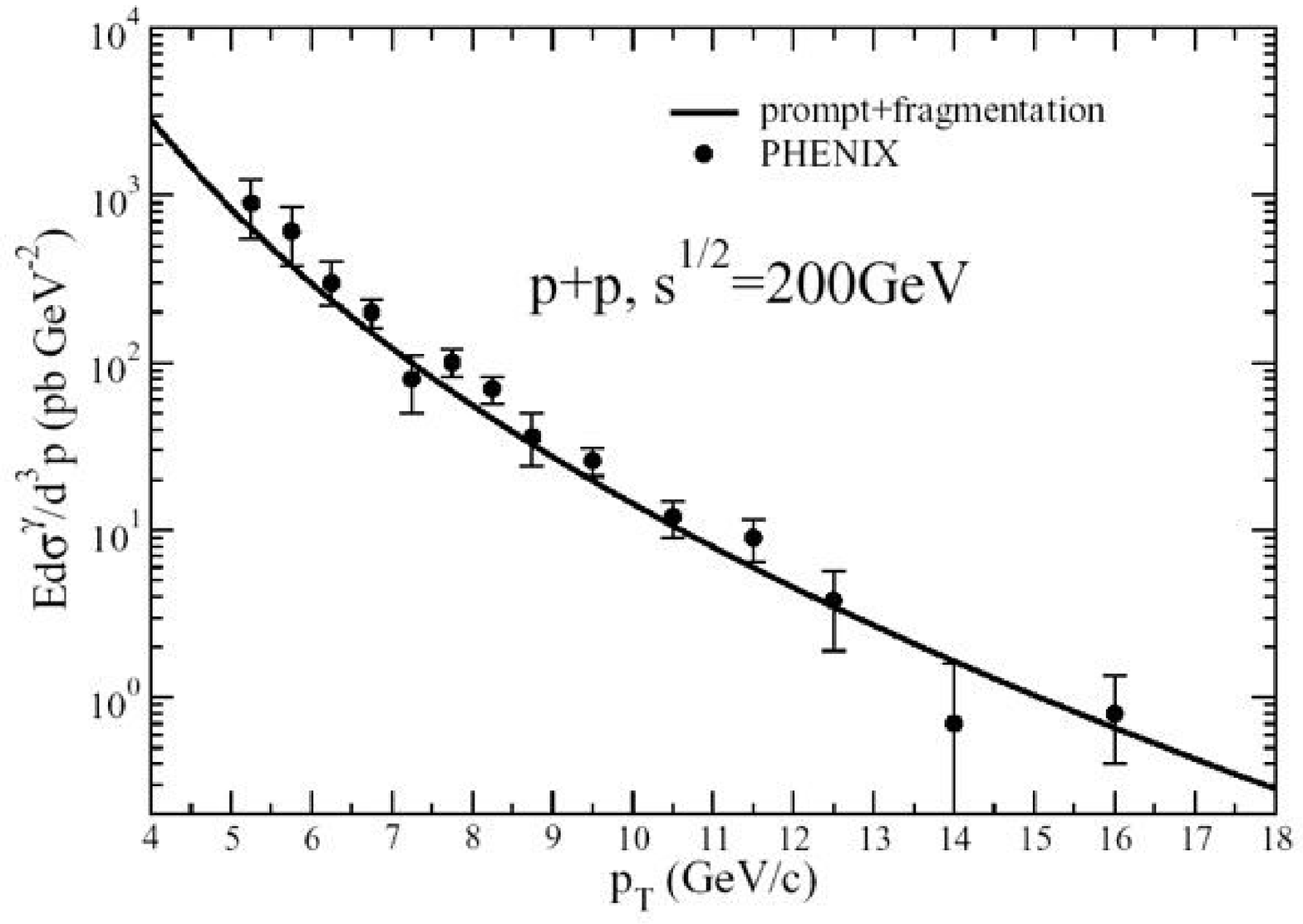}}
\hspace*{2mm}
\resizebox{6.5cm}{!}{\includegraphics{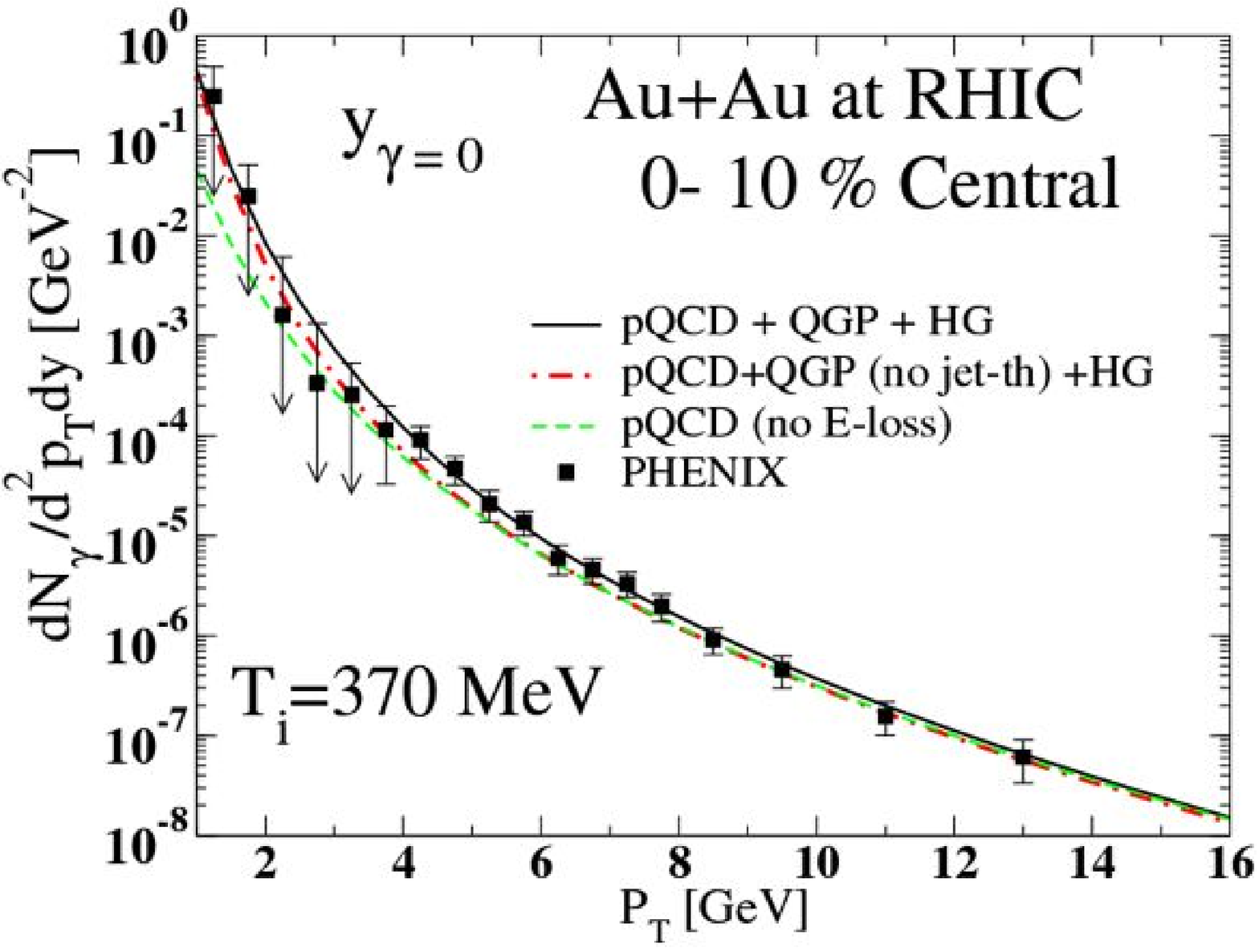}}
\caption{{\sl Left panel:} Photon spectrum from p-p collisions at 200 GeV by
the PHENIX collaboration \cite{PHENIXpp} compared to a perturbative QCD
calculation.  {\sl Right panel:} Comparison of the preliminary photon spectrum
for Au+Au at 200 A GeV from PHENIX \cite{PHENIXgold} to theoretical predictions
\cite{Gale}.}
\end{center}
\end{figure}

The left panel of Fig. 4 shows data from WA98 \cite{wa98} on the production of
direct photons (after subtraction of hadronic decays, mainly $\pi^0$ and $\eta$-
meson) in central collisions of Pb+Pb at the CERN SPS.  In comparison is the
prediction of perturbative QCD for hard scattering; obviously it falls short,
indicating an extra source.  The right panel of Fig. 4 shows the result of
adding thermal radiation to the pQCD prediction.  The dynamical evolution model
is the same as used for dileptons as was shown in the right panel of Fig. 2.
There is very good agreement within the range of measured transverse momenta
from 1 to 4 GeV.

Photons have now been measured at RHIC.  The left panel of Fig. 5 shows pp
measurements by PHENIX at 200 GeV \cite{PHENIXpp} compared to theoretical
calculations of prompt plus fragmentation photons \cite{Aurenche,Gordon,Gale}.
This establishes a nice baseline for Au+Au collisions, shown in the right panel
of Fig. 5.  A nice model fit to the data is obtained with a combination of pQCD,
thermal emission from QCD plasma and hadron gas using boost-invariant Bjorken
hydrodynamics, plus one more new ingredient: conversion of a high energy parton
jet produced early in the nuclear collision via interaction with a thermal quark
or gluon \cite{Fries} (see the paper by Jeon in this volume).  The jet-photon
conversion mediated by the plasma is a new idea and has some interesting
consequences. There is more jet conversion where the medium is thicker, hence
the $v_2$ describing the ellipticity of the photons ought to be negative
\cite{Turbide} (see the talk by Heinz in this volume).  Therefore it ought to be
possible to separate photons produced via this mechanism from the thermal or
prompt photons.  In this sense these photons will be a hard probe of the medium
through which the jets are moving.

{\bf 5 Conclusion}

There are two conclusions I wish to emphasize:\\
$\bullet$ Solid results are being obtained, both theoretically and
experimentally, about many-body physics at high energy density, such as
modification of vector spectral densities and QCD processes at high energy.\\
$\bullet$ It is very important to clearly separate the correlation or response
functions characterizing a system in thermal equilibrium from the space-time
evolution characterizing a heavy ion collision.\\
There is plenty of exciting work ahead of us and much to accomplish!

{\bf Acknowledgements}

This work was supported by the US Department of Energy (DOE) under grant
DE-FG02-87ER40328.

\end{document}